\documentclass{llncs}
\usepackage{amssymb} 
\usepackage{epsfig}

\begin{document}
\pagestyle{plain}  
\bibliographystyle{splncs}

\title{A Constructive Generalization\\of Nash Equilibrium}

\author{Xiaofei Huang}

\institute{eGain Communications, Mountain View, CA 94043 \\
\email{huangxiaofei@ieee.org}}

\maketitle
%

\begin{abstract}
In a society of multiple individuals, 
	if everybody is only interested in maximizing his own payoff, 
	will there exist any equilibrium for the society?
John Nash proved more than 50 years ago that an equilibrium always exists
	such that nobody would benefit from unilaterally changing his strategy.
Nash Equilibrium is a central concept in game theory,
	which offers the mathematical foundation for social science and economy.
However, the original definition is declarative without including a solution to find them.
It has been found later that it is computationally difficult to find a Nash equilibrium.
Furthermore, a Nash equilibrium may be unstable, sensitive to the smallest variation of payoff functions.
Making the situation worse, a society with selfish individuals can have an enormous number of equilibria,
	making it extremely hard to find out the global optimal one.
This paper offers a constructive generalization of Nash equilibrium
	to cover the case when the selfishness of individuals are reduced to lower levels in a controllable way.
It shows that the society has one and only one equilibrium when the selfishness is reduced to a certain level.
When every individual follows the iterative, soft-decision optimization process presented in this paper,
	the society converges to the unique equilibrium with an exponential rate under any initial conditions.
When it is a consensus equilibrium at the same time, it must be the global optimum.
The study of this paper suggests that, to build a good, stable society (including the financial market) for the benefit everyone in it, 
	the pursuing of maximal payoff by each individual should be controlled at some level
	either by voluntary good citizenship or some proper regulations.
\end{abstract}

\section{Introduction}
John Nash has proved in 1950 using Kakutani fixed point theorem that any $n$-player normal-form game~\cite{GameTheoryLuce}
	has at least one equilibrium.
In $n$-player normal-form game, each player has only a finite number of actions to take and 
	takes one strategy at action playing.
If a player takes one of the actions in a deterministic way,
	it is called a pure strategy.
Otherwise, if a player takes anyone of the actions following some probability distribution defined on the actions,
	it is called a mixed strategy.
At a Nash equilibrium,
	each player has chosen a strategy (pure or mixed)
	and no player can benefit by unilaterally changing his or her strategy while the other players keep theirs unchanged. 

Nash Equilibrium is arguably the most important concept in game theory,
	which has significant impacts on many other fields like social science, economy, and computer science.
It is an elegant theory for understanding a very important scenario in game playing.

However, the original definition is not constructive.
It does not offer a solution to find them.
Recent studies found that finding a Nash equilibrium is computationally hard (PPAD-complete)~\cite{Daskalakis05three-playergames,DaskalakisGoldbergPapadimitriou05}
	even for 2-player games~\cite{Chen&Deng06}.
The state of the art of existing computer algorithms are Lemke-Howson~\cite{Lemke&Howson64} for 2-player games,
	Simplicial Subdividison~\cite{Laan87} and Govindan-Wilson~\cite{Govindan03} for N-player games.
	
A Nash equilibrium may not be stable.
A mixed strategy equilibrium is always very sensitive to perturbation and computing errors.
A smallest change in utility function or a slightest round-off error could knock the players
	out their equilibrium with mixed strategies.
Furthermore, a N-player game may have a huge number of Nash equilibria, growing exponentially with the number of players.
The players can be trapped into one equilibrium or another, sensitive to initial conditions and perturbations.
Finding the optimal one turns out to be a NP-hard problem.

Often times, 
	the memory, information exchange,  and computing power are imperfect and limited for real living beings in a society.
We can imagine that it is not an easy task for them to reach a Nash equilibrium.
The Nash equilibrium is defined by selfish individuals trying to maximizing their own payoffs.
Our experiences tell us that a society with selfish individuals
	may not be able to yield good payoffs to everyone in it.
Such a society could be unstable, 
	quickly sways from one state to another, 
	and never be being able to reach an equilibrium.
Could we build a good, efficient, and stable society by simply reducing the selfishness of individuals in a society?

Our conventional wisdom tells us
	that if each of us gives away a bit more in favor of others,
	we could end up with more gains as return.
That is, the reduced selfishness leads to better payoffs for the individuals in a society.
For instance, if we, as drivers, 
	respect other drivers sharing the same road and 
	give considerations for each other either voluntarily and/or by following traffic laws,
	then each of us will end up with a faster, safer drive to destination 
	than the case when everyone is only interested in maximizing his own speed to destination.

This paper offers a constructive generalization of Nash equilibrium
	along the line of reducing selfishness.
It is based on a recently discovered general global optimization method, 
	called cooperative optimization~\cite{Huang03Greece,HuangBookCCO,HuangDAGM04}.
Cooperation is an ubiquitous phenomenon in nature.
The cooperative optimization theory is a mathematical theory 
	for understanding cooperative behaviors and translating it into optimization algorithms.


\section{A Constructive Generalization}	

There is a fundamental difference between cooperative optimization and many classical optimization methods.
It is at the very core of optimization, 
	i.e., the way of making decisions for assigning decision variables.
Classic ones often times make precise decisions at assigning variables 
	at a given time instance of optimization, such as $x=3$ for the time instance $t$.
Such an assignment is precise at the sense that $x$ can only be the value of $3$, not any other ones.
In contrast, the former makes soft decisions, represented by probability like functions called assignment functions, 
	such as $\Psi(x, t)$, at the time instance $t$.
It says that at the time instance $t$, the variable $x$ can be of any value 
	with the likelihood measured by the function value $\Psi(x, t)$.
A variable value of a higher function value is more likely to be assigned as a value to the variable  
	than another value of a lower function value.

If the function $\Psi(x, t)$ at time $t$ is peaked at a specific value, say $x=3$, 
	then the soft decision falls back to the classic precise decision, 
	e.g., assigning the value $3$ to the variable $x$ ($x=3$).
Hence, soft decision making is a generalization of the classic precise decision making.	

Let $E(x_1, x_2, \ldots, x_n)$ (or simply $E(x)$) be a multivariate objective function of $n$ variables. 
Assume that $E(x)$ can be decomposed into $n$ sub-objective functions $E_i(x)$, 
	one for each variable, such that those sub-objective functions satisfying
\[ E_1(x) + E_2(x) + \ldots + E_n(x) = E(x) \ , \]
and/or the maximization of $E_i(x_i)$ with respect to $x_i$ also leads to the maximization of $E(x)$ for any $i$.

In terms of a multi-agent system, 
	let us assign $E_i(x)$ as the objective function for agent $i$, for $i=1,2,\ldots, n$.
There are $n$ agents in the system in total.
The objective of the system is to minimize $E(x)$ and 
	the objective of each agent $i$ is to minimize $E_i(x)$.
In game theory, $E_i(x)$ is called the utility function of agent $i$.
In this paper, $E(x)$ is also called the global utility function of a game.

A simple form of cooperative optimization is defined as an iterative update of
	the assignment function of each agent as follows:
\begin{equation}
\Psi_i (x_i, t) = \sum_{\sim x_i} \left( e^{E_i(x)/\hbar} \prod_{j \not= i} p_j(x_j, t-1) \right),\quad \mbox{for $i=1,2,\ldots,n$}  \ , 
\label{cooperative_optimization_general3}
\end{equation}
where $\sum_{\sim x_i}$ stands for the summation over all variables except $x_i$ and $\hbar$ is a constant of a small positive value.
$p_i(x_i, t)$ is defined as 
\begin{equation} 
p_i(x_i, t) = \left(\Psi_i(x_i,t)\right)^{\alpha} / \sum_{x_i} \left(\Psi_i(x_i,t)\right)^{\alpha} \ , 
\label{compute_assignment_probabilty}
\end{equation}
where $\alpha$ is a parameter of a positive real value.

By the definition, $p_i(x_i, t)$ just likes a probability function satisfying
\[ \sum_{x_i} p_i (x_i, t) = 1 \ . \]
It is, therefore, called the assignment probability function.
It defines the probability-like soft decision at assigning variable $x_i$ at the time instance $t$.

The original assignment function $\Psi_i(x_i, t)$, is called the assignment state function.
That is, the state of agent $i$ at the time instance $t$ is represented 
	by its assignment state function $\Psi_i(x_i, t)$.
From Eq.~\ref{compute_assignment_probabilty} we can see that
	the assignment probability function $p_i(x_i, t)$ is defined 
	as the assignment state function $\Psi_i(x_i)$ to the power $\alpha$ with normalization.
To show the relationship, the assignment probability function $p_i(x_i, t)$ is also expressed as $\left({\bar \Psi}_i(x_i,t)\right)^{\alpha}$ 
	in the following discussions with the bar standing for the normalization.
	
With this notation, the iterative update function~(\ref{cooperative_optimization_general3}) can be rewritten as
\begin{equation}
\Psi_i (x_i, t) = \sum_{\sim x_i} \left( e^{E_i(x)/\hbar} \prod_{j \not= i} \left({\bar \Psi}_j (x_j, t-1)\right)^{\alpha} \right),\quad \mbox{for $i=1,2,\ldots,n$}  \ . 
\label{cooperative_optimization_general3b}
\end{equation}

By substituting Eq.~\ref{cooperative_optimization_general3} into Eq.~\ref{compute_assignment_probabilty},
	we have a mapping from a set of assignment probability functions to itself.
Because the set is compact and the mapping is continuous, 
	so a fixed point exists based on Brouwer fixed point theorem.
Since a set of assignment state functions 
	is uniquely defined by a set of assignment probability functions by Eq.~\ref{cooperative_optimization_general3},
We can conclude that 	
	there exists at least one set of assignment state functions $\{\Psi^{*}_1(x_1),\Psi^{*}_2(x_2), \ldots, \Psi^{*}_n(x_n) \}$ such that
\[\Psi^{*}_i (x_i) = \sum_{\sim x_i} \left( e^{E_i(x)/\hbar} \prod_{j \not= i} \left({\bar \Psi}^{*}_j(x_j)\right)^{\alpha} \right),~~~\mbox{for $i=1,2,\ldots, n$} \ . \]
	
Without loss of generality, let the utility function $u_i(x)$ for the agent $i$ be defined as  
\[ u_i (x) = e^{E_i(x)/\hbar} \ . \]
In this case, the agent $i$ tries to maximize the utility function $u_i(x)$ instead of 
	maximizing the objective function $E_i(x)$ where the former task is equivalent to the latter.
Accordingly, the simple form of cooperative optimization~(\ref{cooperative_optimization_general3b}) becomes
\begin{equation}
\Psi_i (x_i, t) = \sum_{\sim x_i} \left( u_i(x) \prod_{j \not= i} p_j (x_j, t-1)\right),\quad \mbox{for $i=1,2,\ldots,n$}  \ , 
\label{cooperative_optimization_general3a}
\end{equation}
where $p_j(x_j, t)$ is the assignment probability function defined by Eq.~\ref{compute_assignment_probabilty}.

From Eq.~\ref{cooperative_optimization_general3a},
	we can see that the assignment state function $\Psi_i(x_i, t)$ for a given variable value $x_i=a$ is
	the payoff of agent $i$ with the action $a$ (taking only the action labeled by the value $a$) 
	while other players use the mixed strategies $p_j(x_j, t)$ (for the $j$s where $j \not = i$).
An action $a_1$ is better than another action $a_2$ if $\Psi_i(a_1, t) > \Psi_i(a_2, t)$.
The expected payoff of the agent $i$ is determined by the mixed strategy $p_i(x_i, t)$ as follows
\[ \sum_{x_i} \Psi_i(x_i, t)  p_i (x_i, t) \ . \]

The probability assignment function $p_i(x_i, t)$ is also called the strategy of agent $i$ in game theory.
The set of strategies $\{p_1(x_1, t), p_2(x_2, t), \ldots, p_n(x_n, t) \}$
	is called a strategy profile in game theory, denoted as $p$.

The best action of agent $i$ at time $t$ is defined as the one with the highest payoff,
	i.e., the $x_i$ that maximizes $\Psi_i(x_i, t)$.
Assume that the total number of actions of agent $i$ is $m_i$.
Assume further that $\alpha \ge 1$.
Based on its definition given in (\ref{cooperative_optimization_general3a}), 
	we find out the difference between the best payoff $\max_{x_i} \Psi_i(x_i, t)$ 
	and the expected payoff $\sum_{x_i} \Psi_i(x_i, t)  p_i (x_i, t)$.
It is straightforward to derive that the difference should satisfy the following inequality:

\[ 0 \le \max_{x_i} \Psi_i(x_i, t) - \sum_{x_i} \Psi_i(x_i, t)  p_i (x_i, t)< \left(\frac{m_i - 1}{e} \max_{x_i} \Psi_i(x_i, t)\right) \alpha^{-1}\ . \]
Obviously, the difference can be arbitrarily small when the parameter $\alpha$ is sufficiently large.
That is, the difference is reduced to zero when $\alpha \rightarrow \infty$,
\[ \lim_{\alpha \rightarrow \infty} \left( \max_{x_i} \Psi_i(x_i, t) - \sum_{x_i} \Psi_i(x_i, t)  p_i (x_i, t) \right) = 0. \ . \]

Based on Brouwer fixed point theorem, the simple form~(\ref{cooperative_optimization_general3a})
	must also exist an equilibrium (a fixed point) for any $\alpha > 0$.
That is, given any $\alpha > 0$, 
	there exists at least one set of assignment state functions 
	$\{ \Psi^{*}_1(x_1), \Psi^{*}_2(x_2), \ldots, \Psi^{*}_n(x_n) \}$ such that
\[\Psi^{*}_i (x_i) = \sum_{\sim x_i} \left( u_i(x) \prod_{j \not= i} \left({\bar \Psi}^{*}_j (x_j)\right)^{\alpha} \right),
	\quad  \mbox{for $i=1,2,\ldots, n$} \ . \]

At the equilibrium,
	we know from the previous discussion that, for each agent $i$,
	the difference between its best payoff $\max_{x_i} \Psi^{*}_i(x_i)$ 
	and its expected payoff $\sum_{x_i} \Psi^{*}_i(x_i) p^{*}_i (x_i)$ can be arbitrarily small if we choose a sufficiently large parameter $\alpha$.
That is, for any $i$,
\begin{equation}
\lim_{\alpha \rightarrow \infty} \left( \max_{x_i} \Psi^{*}_i(x_i) - \sum_{x_i} \Psi^{*}_i(x_i) p^{*}_i (x_i) \right) = 0 \ . 
\label{cooperation_equilibrium}
\end{equation}

Given a strategy profile $p^{*}$, 
	it is a Nash equilibrium if and only if, 
	given any agent, its best payoff is equal to its expected payoff $\sum_{x_i} \Psi^{*}_i(x_i) p^{*}_i (x_i)$.
That is, for any $i$,
\begin{equation}
\max_{x_i} \Psi^{*}_i(x_i) - \sum_{x_i} \sum_{x_i} \Psi^{*}_i(x_i) p^{*}_i (x_i) = 0 \ . 
\label{Nash_equilibrium}
\end{equation}

Compare the statement~(\ref{cooperation_equilibrium}) with the statement~(\ref{Nash_equilibrium}), 
	we can conclude that
	any equilibrium of the simple form of cooperative optimization~(\ref{cooperative_optimization_general3a}) 
	can be arbitrarily close to a Nash equilibrium if the parameter $\alpha$ is sufficiently large.
The simple form not only offers a general definition of a new kind of equilibria,
	but also provides an algorithmic method for finding them.

A very large value for the parameter $\alpha$ stands for a very selfish agent.
To make this point clear, we can take a look at Eq.~\ref{compute_assignment_probabilty}
	used for computing the strategy of each agent at the time instance $t$.
With a very large value $\alpha$,
	each agent greatly amplifies the probability for its best action(s) that has the best payoff at the time instance $t$.
At the same time, the probabilities of its sub-optimal action(s) that offer less payoffs than the best one are significantly suppressed near to the value zero.
Equivalently, we can say that each agent is selfish because he is only interested in maximizing his own payoff.

This observation explains why a Nash equilibrium may not be stable 
	since it can be extremely sensitive to perturbations and errors introduced by the communications among agents, 
	and variations in utility functions.
For example, a slightest variation in the utility function could lead to a dramatic shift of the equilibrium
	from one point in the strategy profile space to another one.
It will be hard for an algorithmic method to converge to an unstable equilibrium purely based on iterations.
	
As a summary, we can say that pursuing the maximal payoff by every player in a game 
	often lead to the difficulty for the game to reach an equilibrium.
Even if an equilibrium is found, it could also be unstable, very sensitive to small changes in the utility functions.
Furthermore, the final payoff for each player in the game may be good enough.
Can the situation be improved if we simply reduce the selfishness of agents 
	by tuning down the parameter $\alpha$?
	
\section{Towards the Global Optimum}	

It is desirable to define some kind of equilibria that are stable and easy to find.
It would be ideal if there exists one and only one equilibrium for a game 
	and the equilibrium is also the social optimum (the global optimum of the global utility function $E(x)$) at the same time.
Often times, a social optimum of a society leads to better payoffs for individuals in the society.
At least, it is the best on average for each individual,
	realizable through the wealth redistribution at certain degree through some social welfare system.
It will be shown in this section that these are possible 
	if the simple form of cooperative optimization~(\ref{cooperative_optimization_general3})
	is converged back to the original general form of cooperative optimization
	and the value of the parameter $\alpha$ is reduced below a certain threshold.
	
From the iterative update function~(\ref{cooperative_optimization_general3b}) defining the simple form,
	we can replace the constant $\alpha$ by $\lambda (t) w_{ij}$, where both $\lambda(t)$ and $w_{ij}$ are constant parameters.
With that substitution, the equation becomes as follows,
\begin{equation}
\Psi_i (x_i, t) = \sum_{\sim x_i} \left( e^{E_i(x)/\hbar} \prod_{j \not= i} \left({\bar \Psi}_j (x_j, t-1)\right)^{\lambda (t) w_{ij}} \right) \ , 
\label{cooperative_optimization_general3d}
\end{equation}

Further note that a maximization operator can be approximated by a summation operator as follows:
\[ \max_{x} e^{f(x)/\hbar} \approx \sum_{x} e^{f(x)/\hbar} \ . \]
(Under the assumption that the function $f(x)$ has a unique global maximum.)

Such an approximation becomes accurate when $\hbar \rightarrow 0^+$, i.e.,
\[ \lim_{\hbar \rightarrow 0^+} \left( \max_{x} e^{f(x)/\hbar} - \sum_{x} e^{f(x)/\hbar} \right) = 0 \ . \]

With this approximation, the iterative update function~(\ref{cooperative_optimization_general3d}) becomes
\begin{equation}
\Psi_i (x_i, t) = \max_{\sim x_i} \left( e^{E_i(x)/\hbar} \prod_{j \not= i} \left({\bar \Psi}_j (x_j, t-1)\right)^{\lambda (t) w_{ij}} \right) \ . 
\label{cooperative_optimization}
\end{equation}
Taking the logarithm of the both sides, we have the following maximization problem,
\begin{equation}
\Psi_i (x_i, t) = \max_{\sim x_i} \left( E_i(x) + \lambda (t) \sum_{j \not= i} w_{ij} \Psi_j (x_j, t-1) \right), \quad \mbox{for $i=1,2,\ldots, n$} \ . 
\label{cooperative_optimization}
\end{equation}
This is the original general form of cooperative optimization.

In this form, each agent optimizes the compromised utility function defined at the right side of the above equation.
It is called the compromised utility function in the sense that it is the linear combination of the original utility function $E_i(x)$
	for agent $i$ and the assignment state functions $\Psi_j (x_j, t-1)$ of other agents $j$ at the previous time instance $t-1$.
The assignment state function $\Psi_i (x_i, t)$ stores 
	the best payoffs in terms of the compromised utility function
	given different values for variable $x_i$.
Therefore, it is also called the assignment payoff function in the general form.

In summary, the general form of cooperative optimization defines a multi-agent system.
In the system,every agent compromises its own utility function by taking into account the possible payoffs of other agents
	and all agents optimize their own compromised utility functions altogether at the same time in parallel.
Therefore, such a multi-agent system is distributed and autonomous,
	making it highly scalable 
	and less vulnerable than a centralized one to perturbations and disruptions on the agents in the system 

Given an assignment payoff function $\Psi_i(x_i, t)$ of agent $i$ at iteration time instance $t$,
	let $\tilde{x}_i(t)$ be the value of $x_i$ maximizing the function, i.e.,
\begin{equation}
\tilde{x}_i(t) = \arg \max_{x_i} \Psi_i(x_i, t) \ . 
\label{best_assignment}
\end{equation}
It represents the best value of $x_i$ at iteration time instance $t$ 
	that gives the highest payoff.
In other words, assigning $\tilde{x}_i(t)$ to $x_i$ leads to 
	the maximization of the compromised utility function defined at the right side of (\ref{cooperative_optimization}).
The solution of the system at iteration time instance $t$ is the collection of those best values as follows
\[ (\tilde{x}_1(t), \tilde{x}_2(t), \ldots, \tilde{x}_n(t)) \mbox{ or simply as } \tilde{x}(t) \ . \]

The parameters $w_{ij}$ ($1 \le i, j \le n$) in (\ref{cooperative_optimization}) 
	control the propagation of assignment payoff functions $\Psi_j(x_j, t)$ ($j=1,2,\ldots, n$) 
	among the agents in the system.
All of $w_{ij}$s together form a $n \times n$ matrix called the propagation matrix $W$. 
To have $\sum_i E_i(x)$ as the global utility function to be maximized,
	it is required that the propagation matrix $W=(w_{ij})_{n \times n}$ is non-negative and satisfies
\[ \sum^n_{i=1} w_{ij} = 1, \quad \mbox{for $j=1,2,\ldots,n$} \ . \]
	
The propagation matrix $W$ has exactly the same property as a transition matrix at describing a Markov chain.
To have assignment payoff functions $\Psi_j(x_j, t)$ uniformly propagated among all the agents,
	it is required that the propagation matrix $W$ is irreducible and aperiodic.
A matrix $W$ is called reducible if there exists
   a permutation matrix $P$ such that $PWP^T$
   has the block form
\[\left(
    \begin{array}{cc}
      A & B \\
      O & C 
    \end{array}
\right)\ . \]  

Given a constant cooperation strength $\lambda(t)$ of a non-negative value less than 1, i.e., $\lambda(t) = \lambda$ 
	and $0 \le \lambda < 1$ for every time instance $t$,
	the general form of cooperative optimization~(\ref{cooperative_optimization}) has one and only one equilibrium. 
It always converges to the unique equilibrium with an exponential rate 
	regardless of initial conditions and perturbations. 
	
To be more general, assume that the agent $i$'s utility function $E_i(x)$ is defined on variable set $X_i$.
Recall that the solution at iteration $t$ is $\tilde{x}(t)$ (see (\ref{best_assignment})).
Let $\tilde{x}(t)(X_i)$ denote the restriction of the solution on $X_i$.
The solution $\tilde{x}(t)$ is called a consensus solution 
	if it is the optimal solution for each optimization problem defined by (\ref{cooperative_optimization}).
That is, 
\begin{equation}
\tilde{x}(t)(X_i) = \arg \max_{X_i} \left( E_i(x) + \lambda(t) \sum_{j \not=i} w_{ij} \Psi_j(x_j, t-1)\right),~~\mbox{for $i=1,2,\ldots, n$}. 
\label{consensus_solution}
\end{equation}

It is important to note that if the general form of cooperative optimization discovers a consensus solution
	at any time instance $t$, then it must be a pure strategy Nash equilibrium.
This conclusion is obvious from the definition of consensus solution given in Eq.~\ref{consensus_solution},
	where no agent $i$ would get higher payoff from unilaterally changing its best assignment $\tilde{x_i}(t)$.
Furthermore, if it converges to a consensus equilibrium
	with a constant $\lambda$ satisfying $0 \le \lambda < 1$,
	then it is both  a pure strategy Nash equilibrium and 
	the social optimum defined as the global optimum of the global utility function of the game,
\[  E_1(x) + E_2 (x) + \cdots + E_n(x) \ . \]
When a game has an enormous number of Nash equilibria, 
	it is important to find the global optimal one.

\section{Conclusions}

This paper presented a multi-agent system for a constructive generalization of Nash equilibrium.
The dynamics of the system is defined by a general global optimization method, called cooperative optimization.
The selfishness of each agent is defined by a parameter used at computing the agent's strategy during each iteration.
Given any positive value for the parameter, the system always exists an equilibrium.
In particular,
	any equilibrium of the system can be arbitrarily close a Nash equilibrium
	when the parameter controlling the selfishness is sufficiently large.
In this case, each agent in the system is only interested in maximizing its own payoff.

This constructive definition offers an insight into the computational difficulty at finding a Nash equilibrium.
It also offers a perspective from a cooperation point of view at understanding the instability of a Nash equilibrium.
This paper shows that when the selfishness of agents is controlled at some level, 
	better and more stable equilibria could be reached by the system.
Under some proper level, there is only one equilibrium for the system
	and it converges to it at an exponential rate with any initial conditions.
When it is a consensus equilibrium at the same time, 
	it must be the global optimum.

\nocite{Pardalos02}
\nocite{GameTheoryLuce}

\nocite{CPapadimitriou98}


\end{document}